\documentclass[conference]{IEEEtran}
\IEEEoverridecommandlockouts
\usepackage{cite}
\usepackage{amsmath,amssymb,amsfonts}
\usepackage{algorithmic}
\usepackage{graphicx}
\usepackage{textcomp}
\usepackage{xcolor}
\usepackage{hyperref}
\usepackage{caption}
\usepackage{subcaption}
\usepackage{array}

\def\BibTeX{{\rm B\kern-.05em{\sc i\kern-.025em b}\kern-.08em
    T\kern-.1667em\lower.7ex\hbox{E}\kern-.125emX}}
\begin{document}

\title{Ultrasound based Gas Detection: Analyzing Acoustic Impedance for High-Performance and Low-Cost Solutions}

\author{\IEEEauthorblockN{Ayush Singh, Pisharody Harikrishnan Gopalakrishnan and Mahesh Raveendranatha Panicker}
\IEEEauthorblockA{\textit{Dept. of Electrical Engineering} \\
\textit{Indian Institute of Technology Palakkad}\\
122001010@smail.iitpkd.ac.in, hgpisharody@iitpkd.ac.in, mahesh@iitpkd.ac.in}

}

\maketitle

\begin{abstract}
Gas leakage is a serious hazard in a variety of sectors, including industrial, domestic, and gas-powered vehicles. The installation of gas leak detection systems has emerged as a critical preventative measure to solve this problem. Traditional gas sensors, such as electrochemical, infrared point, and Metal Oxide Semiconductor sensors, have been widely used for gas leak detection. However, these sensors have limitations in terms of their adaptation to various gases, as well as their high cost and difficulties in scaling. In this paper, a novel non-contact gas detection technique based on a 40 kHz ultrasonic signal is described. The proposed approach employs the reflections of the emitted ultrasonic wave to detect the gas leaks and also able to identify the exact gas in real-time. To confirm the method's effectiveness, trials were carried out using Hydrogen, Helium, Argon, and Butane gas. The system identified gas flow breaches in 0.01 seconds, whereas the gas identification procedure took 0.8 seconds. An interesting extension of the proposed approach is real-time visualisation of gas flow employing an array of transducers.
\end{abstract}


\section{Introduction}

Gas leak detection has become a critical area of research and development, carrying significant implications for human safety, environmental preservation, and industrial operations. The reliable detection of gas leaks plays a pivotal role in enhancing the integrity of pipeline networks and preventing incidents that can result in substantial physical and financial harm. Conventional gas leak detection methods heavily relied on human senses, such as odour or visual cues. However, the advent of advanced sensor technologies has revolutionised the field of gas leak detection. Among the initial breakthroughs in this domain were chemi-resistive sensors, which employed changes in electrical resistance to detect gas leaks. Early iterations of these sensors utilised metal oxide-based materials, offering improved sensitivity and response time compared to human senses\cite{chemiresistor}. However, they encountered limitations in terms of selectivity and vulnerability to environmental interference\cite{limits_chemiresistive}.

Electrochemical sensors, operating through an electrochemical cell, have also made substantial contributions to gas leak detection. These sensors accurately quantify gas concentrations by measuring the resultant electrical current \cite{gas_review}. Possessing high sensitivity, selectivity, and response time, electrochemical sensors have gained popularity for the detection of various gases. Nevertheless, challenges such as regular calibration and limited lifespan due to electrolyte degradation are associated with these sensors\cite{electrochemical_1}.

Another approach to gas leak detection involves the utilisation  of infrared (IR) sensors, which exploit the absorption properties of gases in the infrared spectrum. By analysing the absorption patterns of emitted infrared radiation, IR sensors can detect and differentiate multiple gases simultaneously. They offer excellent selectivity and are suitable for harsh environmental conditions. However, challenges related to cost, calibration requirements, and limitations in detecting specific gas types are encountered with IR sensors \cite{ir_sensor}.

In this work, a novel, non-contact, high-speed ultrasonic gas leak detection system is presented. The system utilizes a 40 KHz ultrasound signal to detect gas leaks in real time and identify the specific gas type by analyzing the reflected ultrasound. Unlike the methodology described in \cite{gas_conc}, which uses transmitted signal amplitude for gas concentration determination in a pipe, our approach focuses on analyzing the reflected wave to detect gas leaks. The system employs the analysis of both amplitude and phase information of the reflected signal to identify the gas. The methodology described in \cite{gas_conc,heart_beat} is utilized to obtain the amplitude and phase properties. The proposed system is claimed to offer cost reductions compared to existing sensor systems while delivering superior performance in detecting gas leaks in a network of gas pipes with multiple gases. This article addresses the hardware components of the system, the signal handling methods employed, and the measurement approach of the sensor.

\begin{figure}[!t]\centering
	\includegraphics[width=8.5cm]{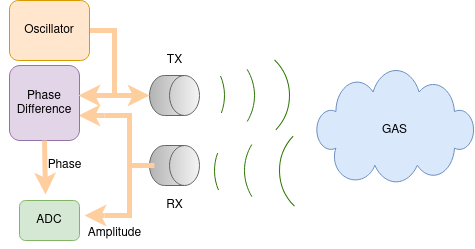}
	\caption{High-level sketch of  architecture of short-range US system with gas detection application. TX: transmitting sensor, RX: receiving sensor, Phase difference:phase difference evaluation block, ADC: analog-to-digital converter, adapted from \cite{heart_beat}}
  \label{fig:high_level}
\end{figure}

\begin{figure*}[ht]\centering
	\includegraphics[width=\linewidth]{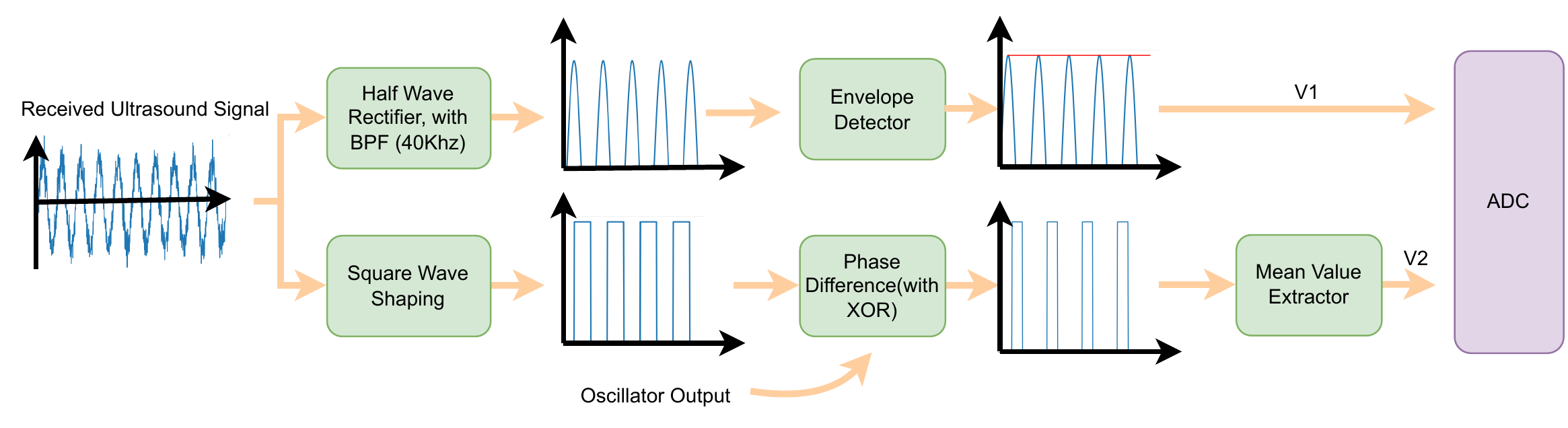}
	\caption{Signal flow overview for the proposed gas sensing unit. Signal $V_1$ represents the amplitude data, whereas signal $V_2$ represents phase change as depicted from \cite{gas_conc,heart_beat}.}
  \label{fig:flow}
\end{figure*}

\section{Methods and Materials}
\subsection{Principle of Detection }
Acoustic impedance is crucial to understanding how sound waves behave when they encounter various materials. It is defined as the measurement of resistance offered by a system to the passage of acoustic waves produced by induced acoustic pressure\cite{acoustic}. The speed at which sound travels through a material is determined by its acoustic impedance; a greater acoustic impedance equates to quicker sound transmission.

When sound waves encounter a boundary between two media of different acoustic impedance, they undergo reflection, with the proportion of the wave reflected dependent on the characteristics of the media or interface\cite{intensity_coeff}. The intensity reflection coefficient quantifies the strength of the reflected wave, calculated as the ratio of the reflected wave's intensity to the incident wave's intensity, as shown in (\ref{reflection coefficient}) \cite{coeff_formula}:

\begin{equation}
    R= \frac{(Z_1-Z_2)^2}{(Z_1+Z_2)^2} = \frac{I_r}{I_i}
    \label{reflection coefficient}
\end{equation}       
where $Z_1$ and $Z_2$ represent the acoustic impedance of the two involved media, $I_r$ is the reflected wave intensity and $I_i$ is transmitted wave intensity. It is important to note that the intensity of the reflected wave is directly proportional to the square of its amplitude, indicating that a larger differential in acoustic impedance results in a higher intensity and amplitude of the reflected wave.

At ambient temperature, acoustic impedance of air is 415$kgm^{-2}s^{-1}$\cite{ac_of_air}, which is lower compared to other materials \cite{z_of_air}. Unless additional gases are present, sound waves are nearly totally reflected back when they meet any substance other than air. However, acoustic impedance of other gases are equivalent to air but not identical, resulting in differing reflected wave amplitudes. The amplitude of the reflected wave can be used to compute the intensity reflection coefficient, and  (\ref{reflection coefficient}) can estimate the acoustic impedance of the spilt gas by comparing it to that of the air. The phase difference of the gas remains significant. A reflected sound wave changes the phase by 180 degrees as it moves from a high to a low acoustic impedance zone \cite{phase_change1,phase_change2}. Furthermore, acoustic impedance is frequently employed to calculate gas density, as in (\ref{acoustic impedance}) \cite{acoustic}.

\begin{equation}
    Z = \rho \cdot C
    \label{acoustic impedance}
\end{equation}
where $Z$ represents the acoustic impedance, $\rho$ denotes the gas density, and $C$ signifies the speed of sound.

\subsection{System Design}
\begin{figure}
    \centering
    \includegraphics[width=8cm,height=7cm]{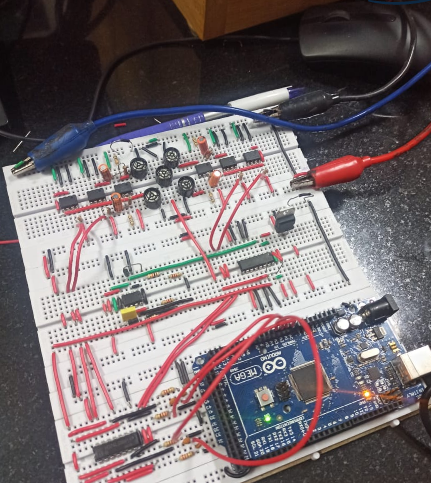}
    \caption{Picture depicting the prototype circuit.}
    \label{fig:circuit}
\end{figure}

\begin{figure}[!b]
    \centering
    \includegraphics[width=4cm, height=5cm]{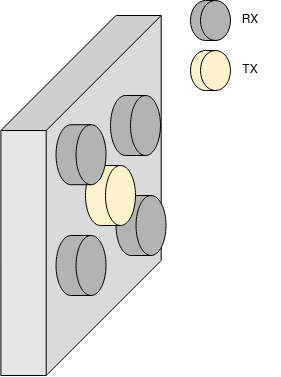}
    \caption{Sensor Placement. TX: transmitting sensor, RX: receiving sensor}
    \label{fig:sensor}
\end{figure}

The proof-of-concept of a remote gas monitoring based on ultrasonic signals is described in this section, which consists of 4 ultrasonic receivers and 1 transmitter. The high-level configuration of the system is as shown in Fig.\ref{fig:high_level}. The transmitter is driven by a 40 kHz square wave, allowing continuous monitoring. The wavelength is 8 mm for a sound speed of 343 m/s. The device uses four receiving transducers in separate positions to reduce noise and ensure high precision as in \cite{heart_beat}. The final measurement is calculated by averaging the four receiver readings. The system utilizes muRATA Piezoelectric Ultrasonic transducers, model MA40S4S\cite{murata}, for both the transmitter and the four receivers. These transducers has a resonant frequency of 40 kHz and a bandwidth of approximately 2 kHz \cite{murata}. The circuit setup with the transmitter and four receivers is depicted in Fig.\ref{fig:circuit}, while Fig.\ref{fig:sensor} illustrates the positioning of the sensors. This configuration and arrangement of components forms the basis of our gas detection system, allowing for the acquisition and processing of relevant signals for gas leak detection.

A voltage-controlled oscillator (VCO) generates a 40 kHz square wave, which is employed to drive the ultrasound transmitter. The reflected signal is received by each of the receiver. Both the amplitude and the phase of the received signal are extracted as illustrated in Fig. \ref{fig:flow}. The filtered signal from a band-pass filter applied to received signal provides the amplitude, denoted as $V_1$. To obtain $V_1$, half-wave rectification and envelope detection techniques are employed. For phase extraction, the incoming signal is converted into square-wave signal ranging from 0 to 5 volts. The phase difference between the transmitted square wave and the converted square wave is determined using XOR logic. This technique generates pulses representing the phase difference, which is passed through an integrator to calculate the mean signal value, $V_2$, which is proportional to the phase difference. The amplitude and phase values are digitized for further analysis by employing an Arduino Mega board with a 10-bit analog-to-digital converter (ADC). The default clock frequency of Arduino is 9600 Hz\cite{arduino1,arduino2}. Adjusting the prescaler to $16$ boosted the sample rate to 80 kHz to meet the Nyquist criterion. The digitized signal is read by Matlab for further processing and analysis.


\subsection{Digital Signal Processing}
The digital signal processing steps consist of Filtering, detection and identification of gases. The first step of filtering is to remove noise and artefacts from the data. The high-frequency components, mostly noise, are filtered out using a modified Gaussian filtering window available in MATLAB. Analysing the amplitude and phase difference in the second step identifies the gas.

In the methodology for gas detection, both the amplitude and phase characteristics of the reflected signal are analysed. The primary objective is to identify significant variations in the amplitude, $V_1$, as it is a reliable indicator of gas leakage. We choose a threshold of $0.1$ V and any increase in amplitude above the threshold is an indicator of gas leakage. Accurately identifying the gas requires precise determination of the reflected signal amplitude, as amplitude signals begin to vary over time as the gas starts diffusing. To achieve this, the first step is to locate the amplitude maxima, which occur at the moment the gas leak occurs, and serve as reliable indicators of the reflected signal amplitude. By calculating the average value based on range of observations around the maxima, we obtain the amplitude of the reflected signal. Utilizing this information, we compute the intensity reflection coefficient, enabling us to determine the acoustic impedance of the spilled gas and identify its composition. Additionally, the phase difference between the incident and reflected signals provides valuable support for our computational analysis.

The proposed approach takes only 0.01 seconds for gas detection and about 0.8 seconds for gas identification on a machine with 16 GB RAM, and AMD Ryzen 5 processor.

\section{Experimental Results}
A mass flow controller (MFC) was used in the studies to perform the experiments. A tube with small diameter was attached to the MFC, and the aperture of the tube was first fixed, then moved to varied distances ranging from 2 cm to 30 cm. The goal was to test the system's repeatability across a range of distances. The studies were conducted using 99.9\% pure gases under typical circumstances of room temperature (25$^o$C) and 50\% humidity. When the distance between the sensor and the tube entrance was 30 cm, the ultrasonic attenuation in air had an influence of roughly 0.02 Volts. Given the comparatively small ultrasonic attenuation at this distance, its influence on the future trials was ignored.

\begin{figure}[ht]\centering
\begin{subfigure}[ht]{\linewidth}
         \centering
         \includegraphics[width=7cm]{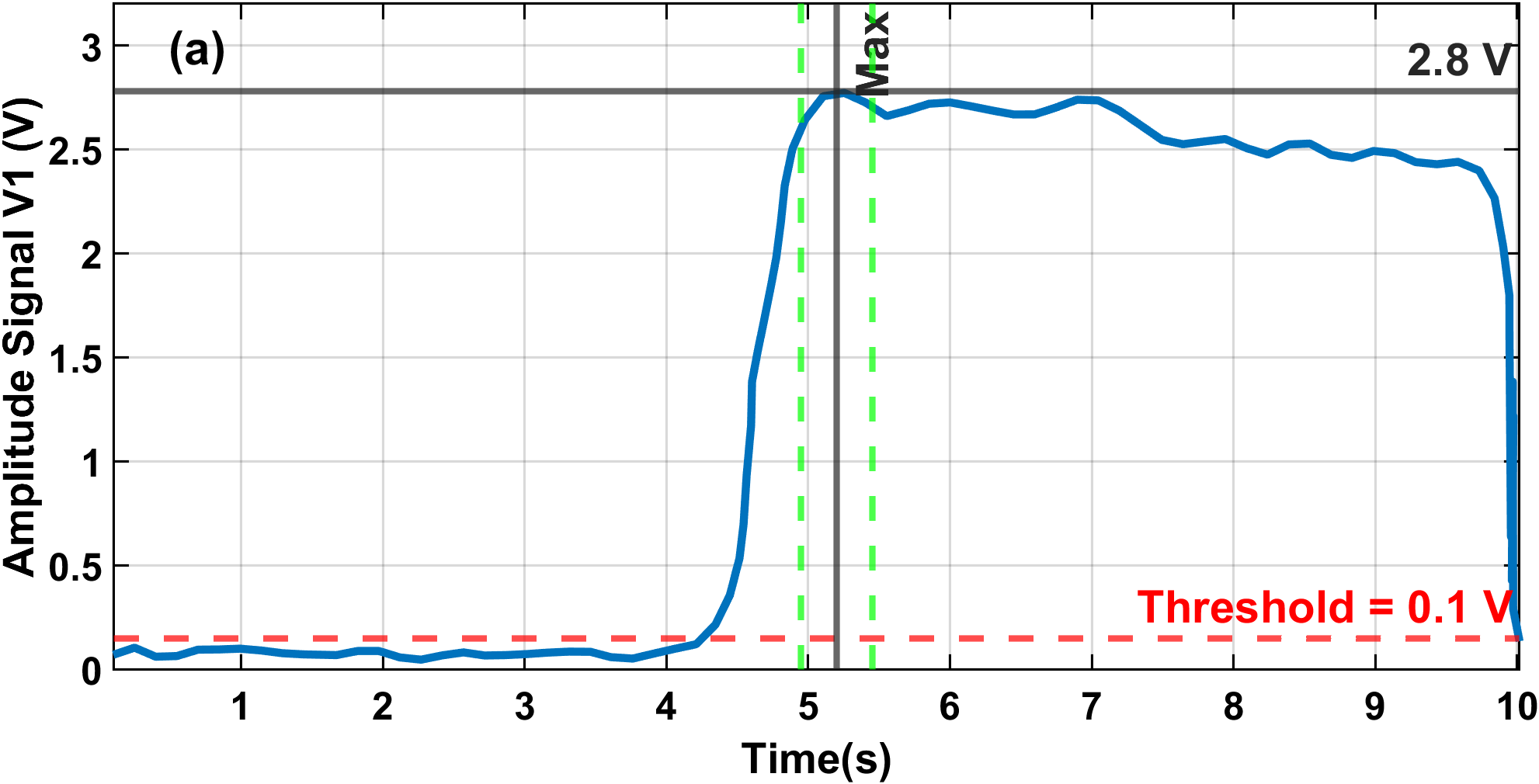}

     \end{subfigure}
     \begin{subfigure}[ht]{\linewidth}
         \centering
        \includegraphics[width=7cm]{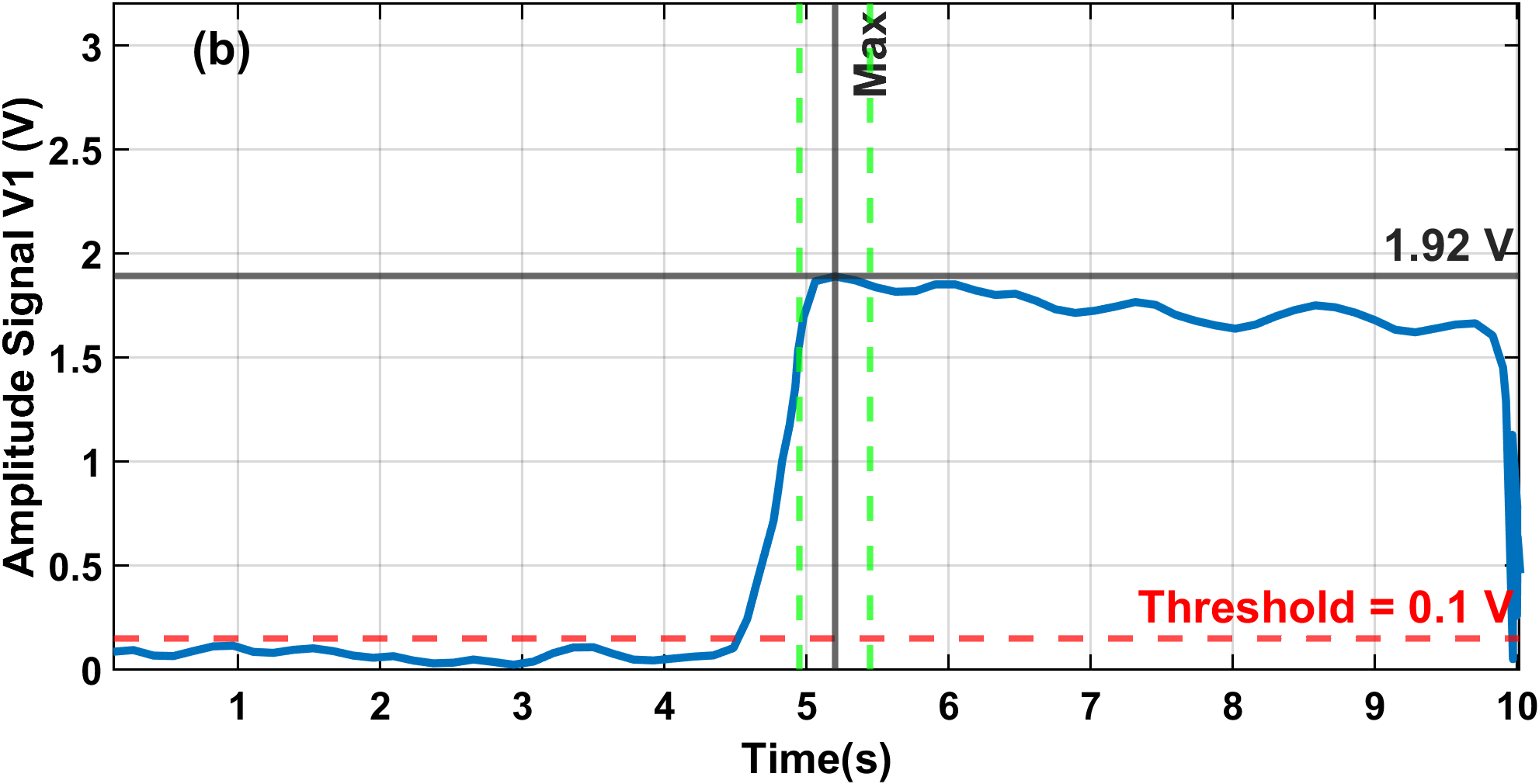}

     \end{subfigure}
          \begin{subfigure}[ht]{\linewidth}
         \centering
        \includegraphics[width=7cm]{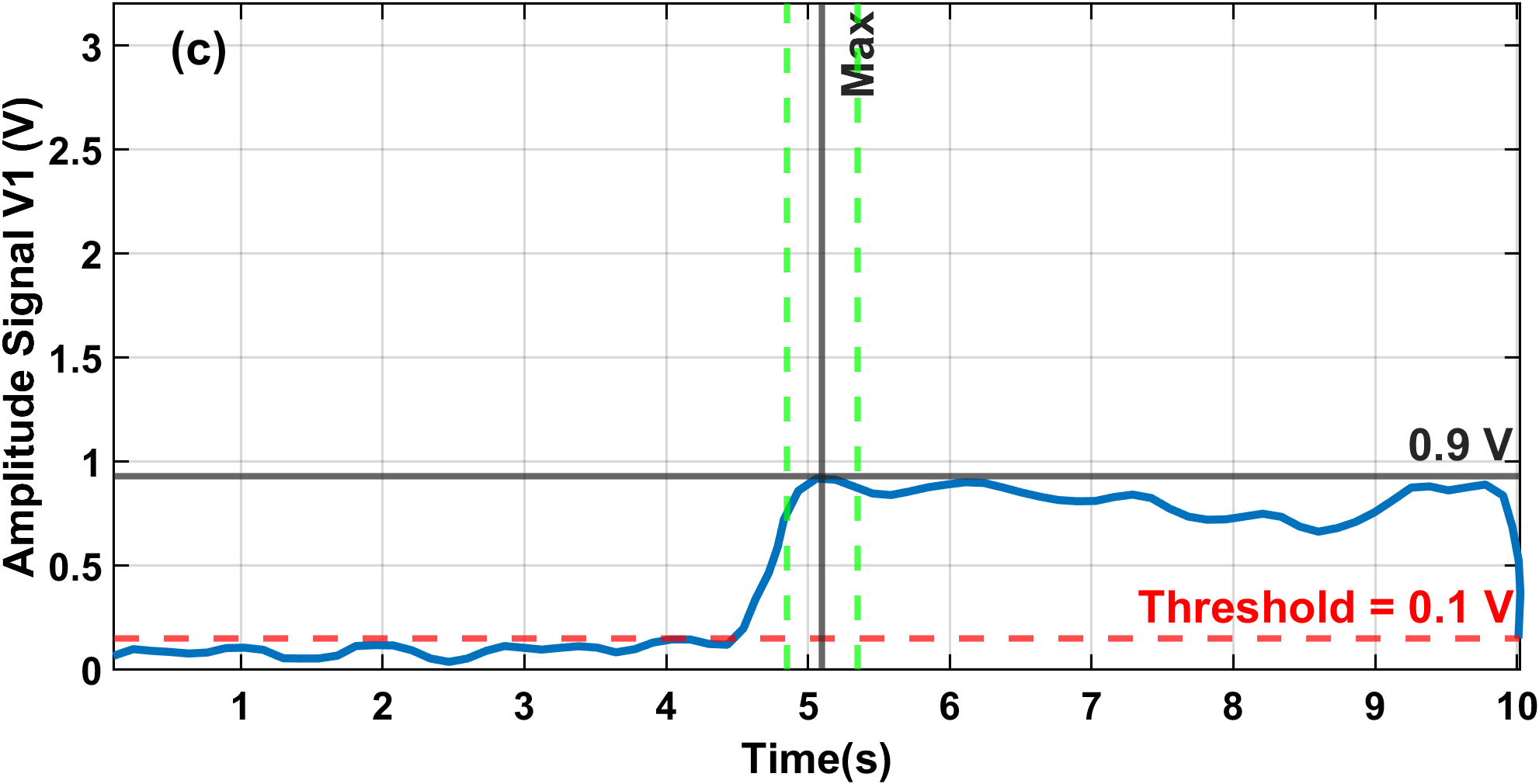}

     \end{subfigure}
          \begin{subfigure}[ht]{\linewidth}
         \centering
        \includegraphics[width=7cm]{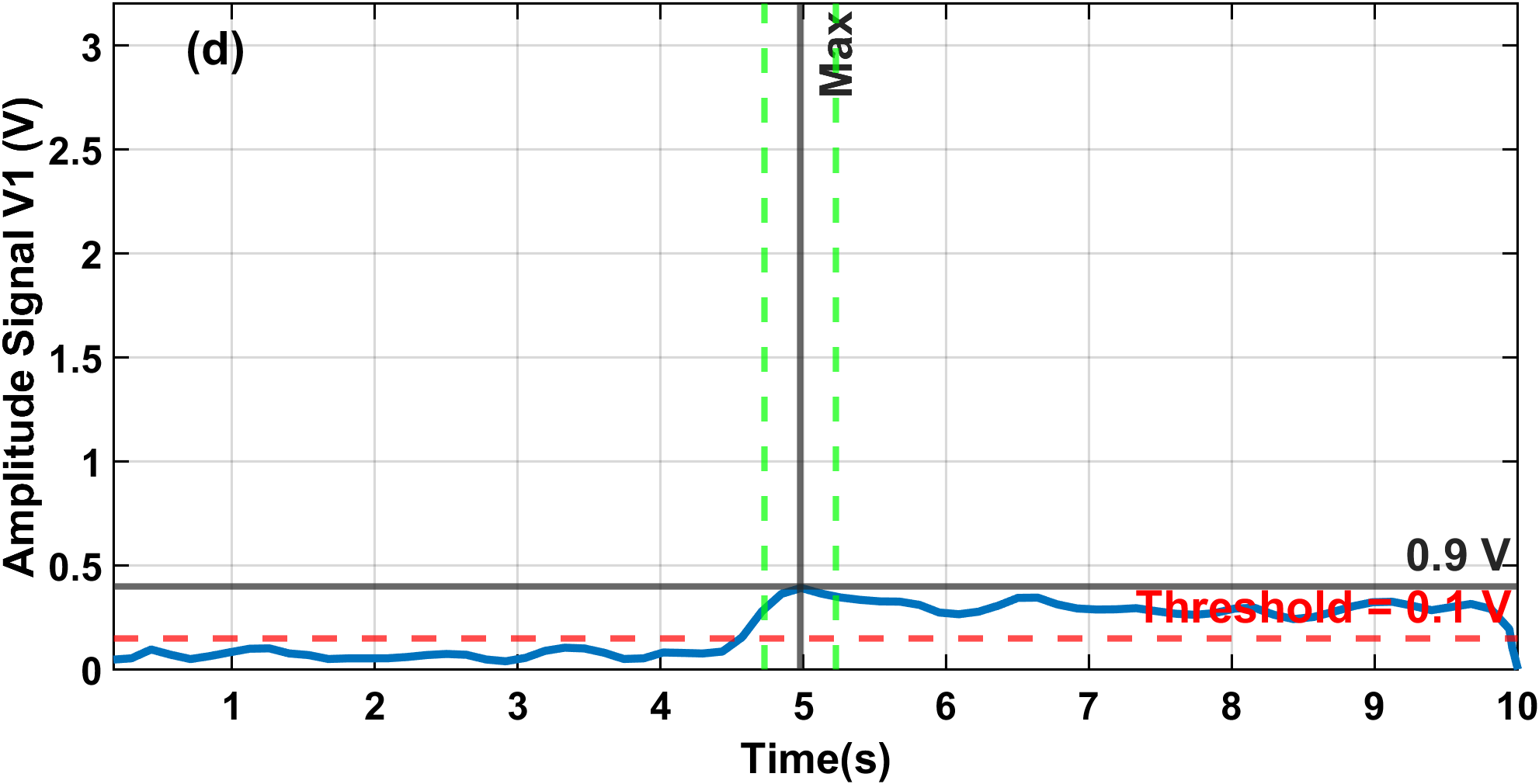}

     \end{subfigure}
	\caption{Amplitude Plot (Signal $V_1$) for (a) Hydrogen (b) Helium (c) Argon (d) Butane}
 \label{fig:gas_identity_amplitude}
\end{figure}

\begin{figure}[ht]\centering
\begin{subfigure}[ht]{\linewidth}
         \centering
         \includegraphics[width=7cm]{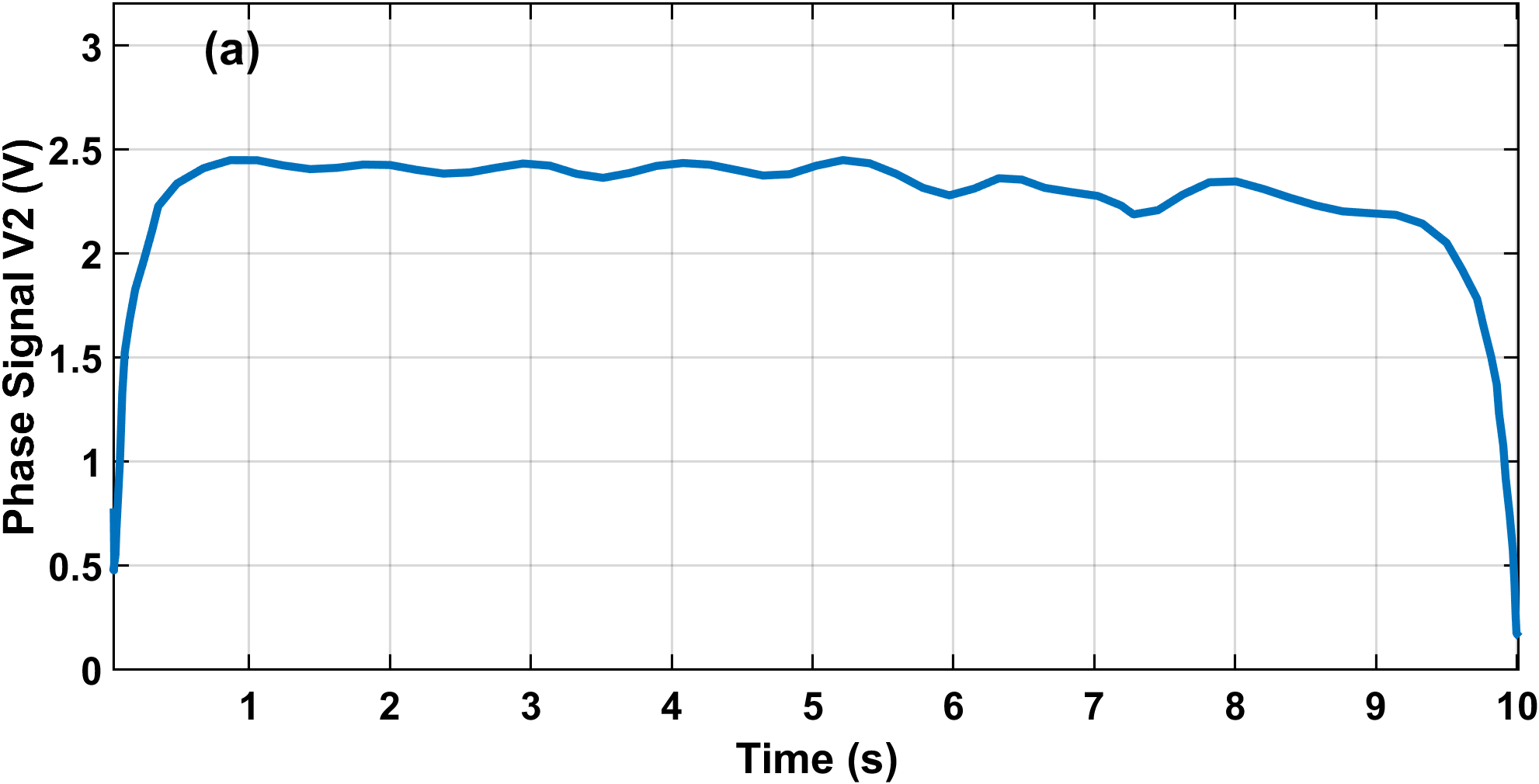}
     \end{subfigure}
     \begin{subfigure}[ht]{\linewidth}
         \centering
        \includegraphics[width=7cm]{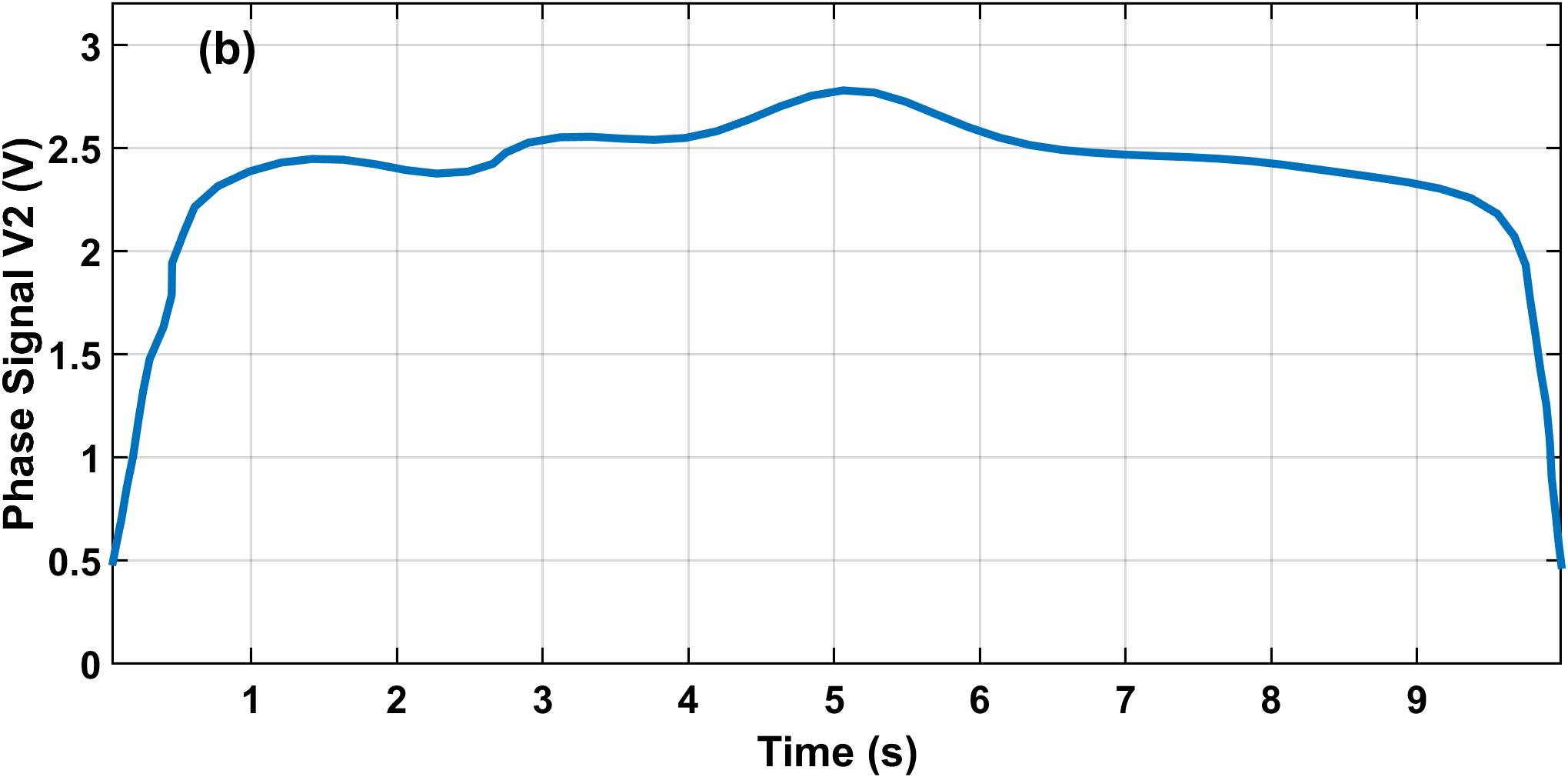}
         
     \end{subfigure}
          \begin{subfigure}[ht]{\linewidth}
         \centering
        \includegraphics[width=7cm]{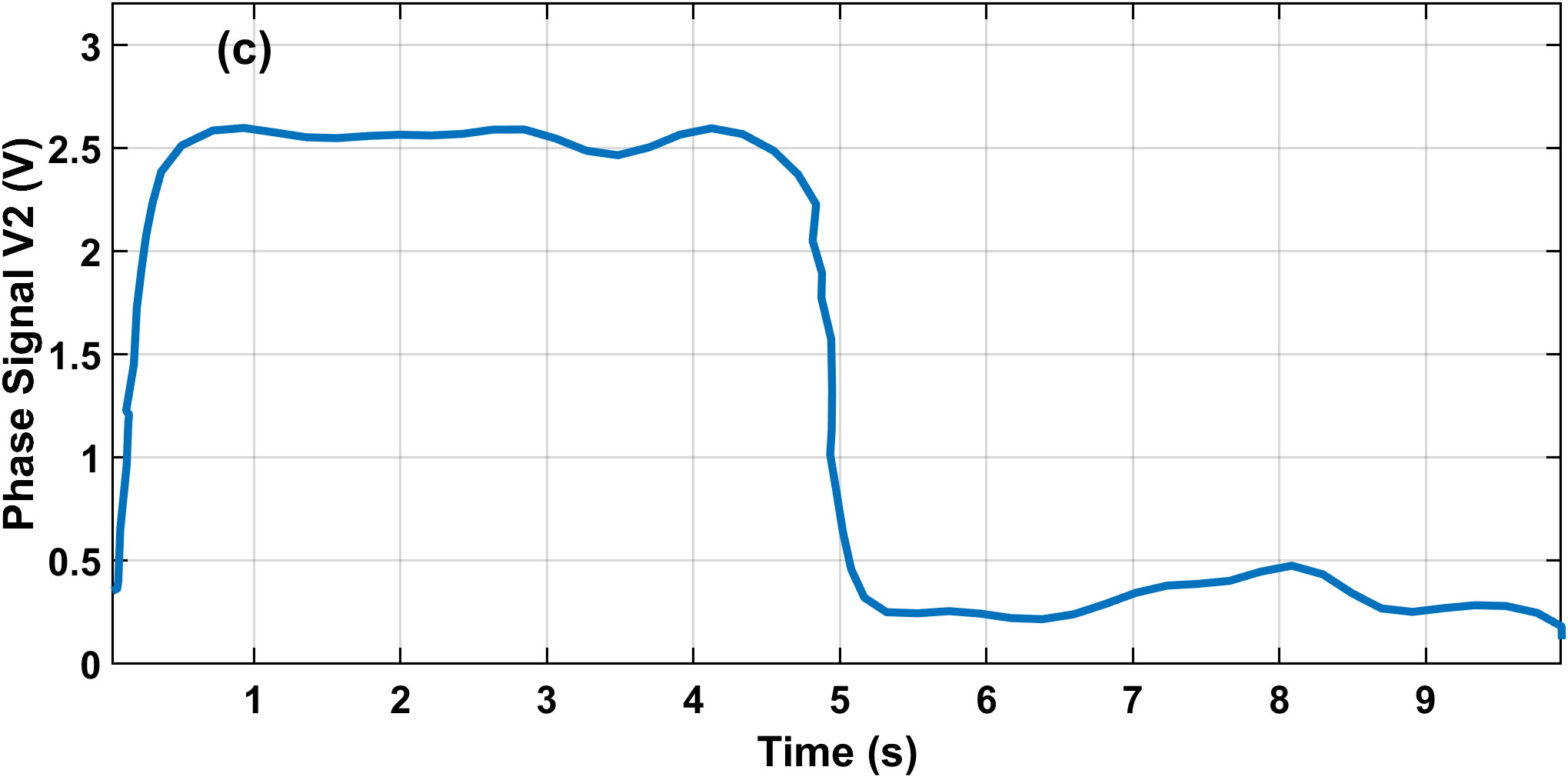}
         
     \end{subfigure}
          \begin{subfigure}[ht]{\linewidth}
         \centering
        \includegraphics[width=7cm]{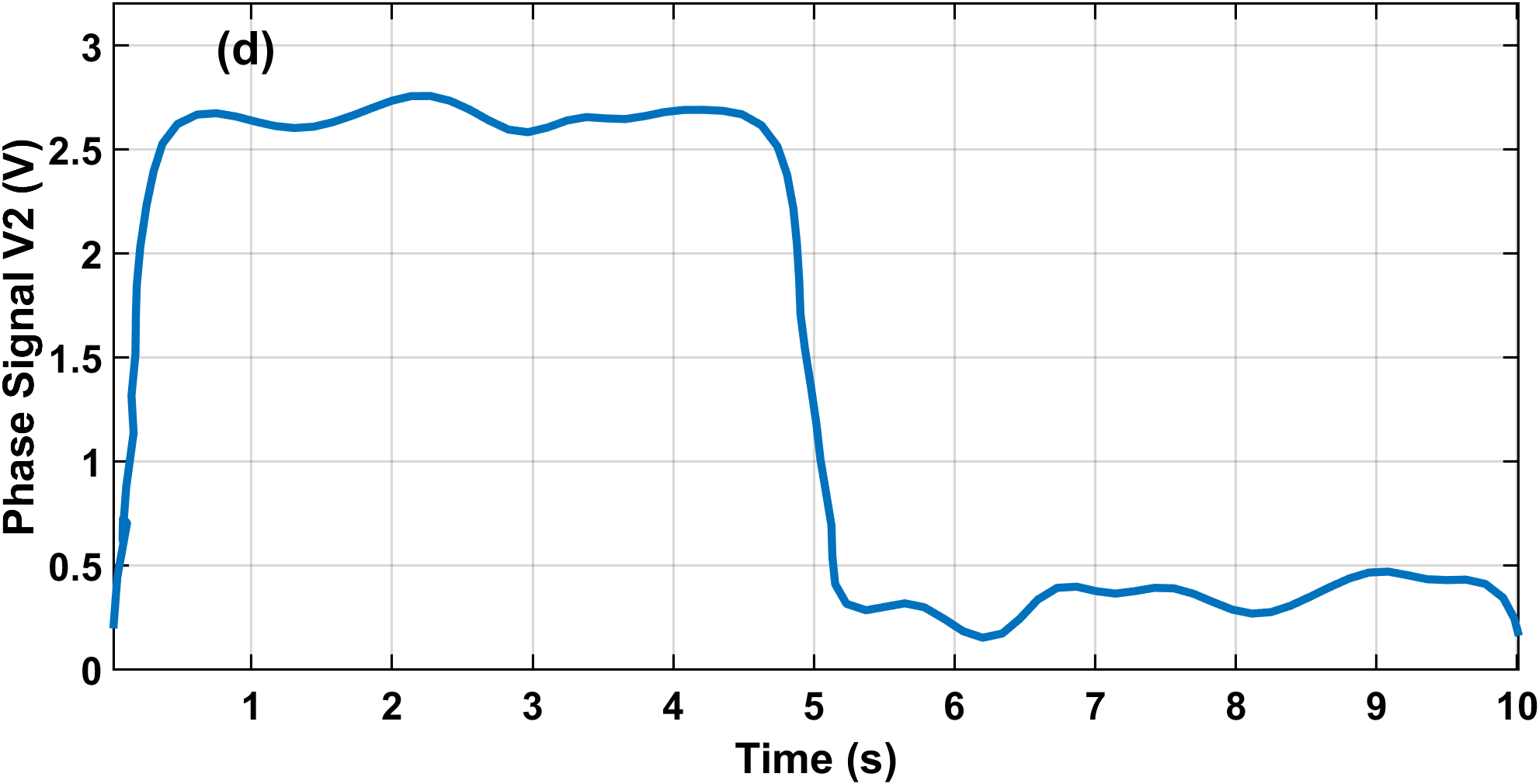}
         
     \end{subfigure}
	\caption{Phase Plot (Signal $V_2$) for (a) Hydrogen (b) Helium (c) Argon (d) Butane}
 \label{fig:gas_identity_phase}
\end{figure}

During the current investigation, gas leakage was deliberately generated without interfering with the passage of gas through the pipeline. During a 10-second observation interval, the system recorded the amplitude and phase signals associated with various gases. The sensor's distance from the pipe's orifice was altered between 2 cm and 30 cm, and the results were recorded. Fig.\ref{fig:gas_identity_amplitude} depicts the amplitude and Fig.\ref{fig:gas_identity_phase} depicts the phase signals estimated by the system across a distance of 15 cm.


From Figures \ref{fig:gas_identity_amplitude} and \ref{fig:gas_identity_phase}, it can be observed that the gas leakage has resulted in changes in the amplitude and the phase of the signals. For the amplitude case, the initial amplitude before the leakage has values around zero, but the values abruptly increased with leakage. It appears that the amplitude sensitivity is most for Hydrogen and least for Butane. However, the phase sensitivity is very good for Argon and Butane when compared to Hydrogen and Helium. Thus by combining both the amplitude and phase values, a very robust system can be developed, which confirms the hypothesis in Section 2. The experiment was repeated with different sensor-pipe distances and that confirmed the trend in the amplitude and the phase difference.    

\begin{figure}[htbp]\centering
	\includegraphics[width=\linewidth]{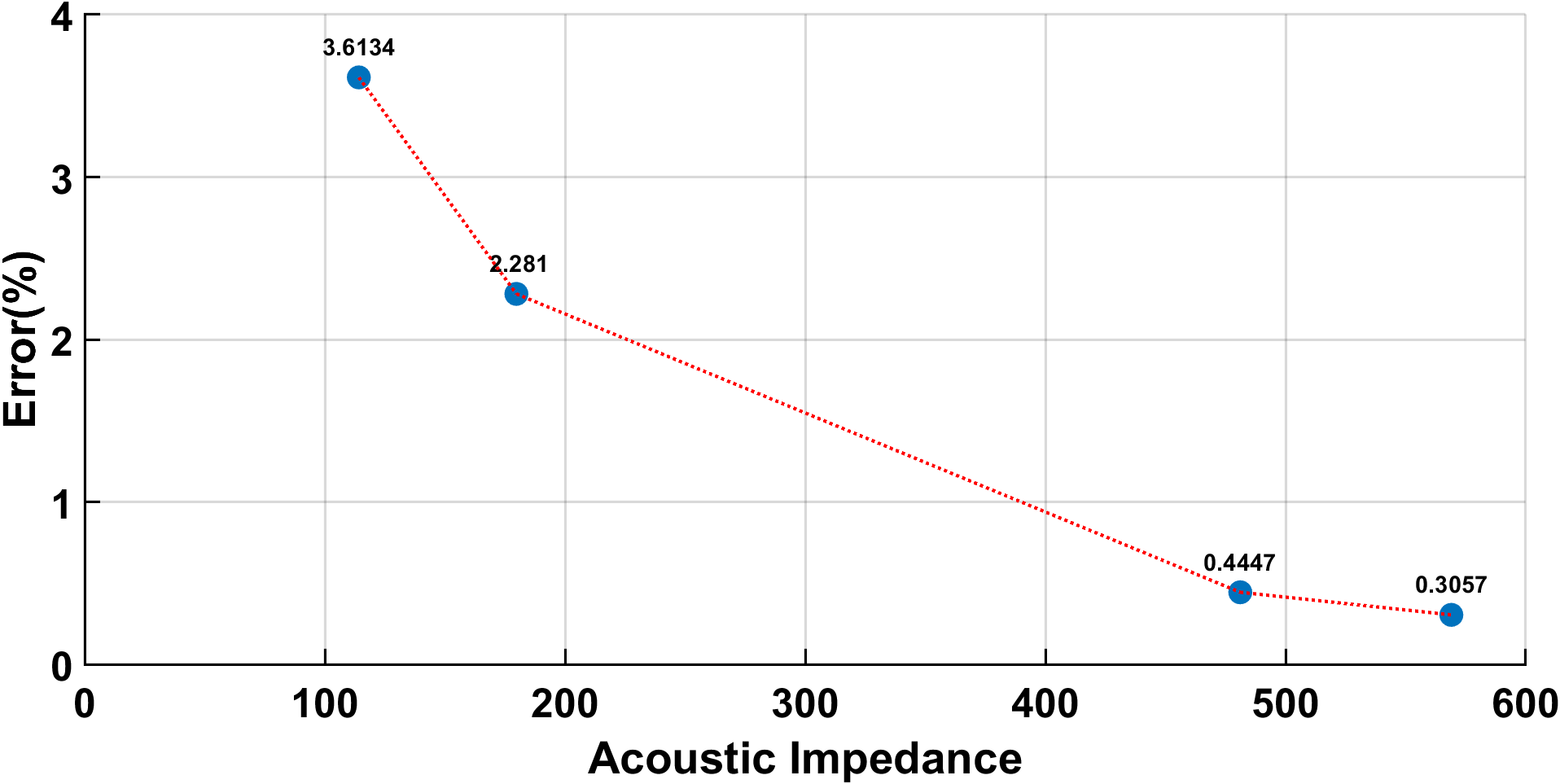}
	\caption{Error plot of gas sensor for Acoustic Impedance}
 \label{fig:error}
\end{figure}

Acoustic impedance was tested at 2–30 cm by averaging the reflected signals from the four receivers. Three readings were collected at each distance between 2cm and 30cm to verify accuracy. Table \ref{table:actual} shows the error in measurement of acoustic impedance. Low error was observed for gases with high acoustic impedance as seen in Fig.\ref{fig:error}

\begin{table}[]
\centering
\caption{Acoustic Properties of Gases}
\begin{tabular}{|m{2cm}|l|l|l|l|}
\hline
\textbf{Gas}                                      & \textbf{Hydrogen} & \textbf{Helium} & \textbf{Argon} & \textbf{Butane} \\ \hline
\textbf{Speed of sound ($m/s$)}                     & 1270\textsuperscript{\cite{speed_of_sound1}}             & 1007\textsuperscript{\cite{speed_of_sound1}}            & 319\textsuperscript{\cite{speed_of_sound1}}            & 194\textsuperscript{\cite{speed_of_sound2}}             \\ \hline
\textbf{Density ($kg/m^3$)}                          & 0.08988\textsuperscript{\cite{density_1}}           & 0.1785\textsuperscript{\cite{density_1}}          & 1.7837\textsuperscript{\cite{density_1}}         & 2.48\textsuperscript{\cite{density_2}}            \\ \hline
\textbf{Actual Acoustic Impedance ($kgm^{-2}s^{-1}$) (\ref{acoustic impedance}) }  & 114.14            & 179.74          & 569.00         & 481.12          \\ \hline
\textbf{Avg. Amplitude $V_1$ (V)}                    & 2.78              & 1.93            & 0.79           & 0.38            \\ \hline
\textbf{Calc. Acoustic Impedance from $V_1$ ($kgm^{-2}s^{-1}$)} (\ref{reflection coefficient}) & 118.419           & 183.84          & 570.74         & 483.26          \\ \hline

\end{tabular}
\label{table:actual}
\end{table}





\section{Conclusion}
In this work, a novel approach to detect and identify the leakage of gases using ultrasonic signal is presented. The approach consisted of one transmitter and four receivers and the amplitude and the phase difference are further employed to detect and identify various gases such as Hydrogen, Helium, Argon and Butane. The Butane gas experimentation produced encouraging findings with high detection rates. However, the performance of the approach under varying conditions of temperature, pressure, and humidity, needs to be investigated further. Future research will also concentrate on improving the system's resilience to external interference and gas diffusion, as well as increasing the detecting distance from the pipe. The suggested approach can also be used to determine a gas's diffusion rate. Another interesting direction will be to employ the approach for visualizing the gas flow employing an array of transducers.


\section*{Acknowledgment}
The authors would like to acknowledge the Central Instrumentation Facility at IIT Palakkad for providing the support during the data acquisition process.

\end{document}